\documentclass[10pt]{iopart}

\expandafter\let\csname equation*\endcsname\relax
\expandafter\let\csname endequation*\endcsname\relax

\usepackage[english]{babel}

\usepackage{cite}

\usepackage{amsmath}
\usepackage{amsthm}
\usepackage{tensor}
\usepackage{amssymb}
\usepackage{bm}
\usepackage{graphicx}
\usepackage[colorlinks, citecolor = blue, urlcolor = blue]{hyperref}

\begin{document}

\title{Spin Hall effects in the sky}

\author{Lars Andersson$^1$, Marius A. Oancea$^2$}
\address{$^1$ Yanqi Lake Beijing Institute of Mathematical Sciences and Applications, Beijing 101408, China}
\address{$^2$ University of Vienna, Faculty of Physics, Boltzmanngasse 5, 1090 Vienna, Austria}
\eads{lars.andersson@bimsa.cn, marius.oancea@univie.ac.at}

\begin{abstract}
In many areas of physics, the propagation of wave packets carrying intrinsic angular momentum is generally influenced by spin-orbit interactions. This is the main mechanism behind spin Hall effects, which result in wave packets following spin-dependent trajectories. Spin Hall effects have been observed in several experiments for electrons in condensed matter systems and for light propagating in inhomogeneous optical media. Similar effects have also been predicted for wave packets propagating in inhomogeneous gravitational fields. We give a brief introduction to gravitational spin Hall effects, emphasizing the analogies with the spin Hall effect of light in optics. Furthermore, we review the most promising astrophysical avenues that could lead to experimental observations of the gravitational spin Hall effect.
\end{abstract}

\section{Introduction}

Hall effects are well known in many areas of physics and represent the basis of many applications of practical interest~\cite{Chien80,Jungwirth2012,SHEL_review}. The most basic example is the ordinary Hall effect, discovered by Edwin Herbert Hall in 1879~\cite{Hall1879} (see Ref.~\cite[Ch. 1]{Ashcroft76} for a standard textbook treatment). This effect is observed for charged particles traveling in a conductor when a magnetic field is applied in the transverse direction. Due to the Lorentz force, charged particles are deflected in the direction orthogonal to the electric current and the applied magnetic field. The magnitude of the deflection is determined by the magnitude of the applied magnetic field and the absolute value of the charge, while the direction of the deflection is given by the orientation of the applied magnetic field and the sign of the charge. 

The ordinary Hall effect, despite its simplicity, successfully captures the main features of Hall effects in general. That is, a particle with some internal degree of freedom traveling in the $x$ direction, under the influence of some external agent acting in the $y$ direction, is deflected in the $z$ direction, and the orientation and magnitude of the deflection depend on the internal degree of freedom. In the case of the ordinary Hall effect, the internal degree of freedom is represented by the charge of the particle. Thus, Hall effects can be viewed as a consequence of the coupling between the external and internal degrees of freedom of particles or wave packets. In this context and throughout this paper, particles should always be understood as an approximate and effective model for the average motion of a localized wave packet.  

Here, we are mainly interested in spin Hall effects. These represent a subclass of Hall effects where the relevant internal degree of freedom is the spin of a particle or wave packet. The main mechanism behind spin Hall effects is the spin-orbit interaction~\cite{SHE_review, SHE_review1, SOI_review}, where the spin represents the internal degree of freedom, and the orbital part represents external degrees of freedom, such as the average position and the average velocity. Due to the mutual interaction between the spin and the external degrees of freedom, spin Hall effects are generally represented by particles or wave packets following spin-dependent trajectories. These effects have been intensively studied for the past 40 years, leading to a solid theoretical description and many experimental observations. The spin Hall effect of electrons is generally present in semiconductors and other condensed matter systems~\cite{originalSHE1,originalSHE2,originalSHE3,originalSHE4,SHE_review,SHE_review1}, while the spin Hall effect of light can be observed for electromagnetic waves propagating in inhomogeneous optical systems~\cite{SHE-L_original,SHE_original,Duval2006,Hosten2008,Bliokh2008,SOI_review,SHEL_review}. Beyond their primary role in fundamental physics, spin Hall effects are also important for many applications in metrology~\cite{SHEL_metrology}, spintronics~\cite{Wunderlich1801,Jungwirth2012}, photonics~\cite{liu2017photonic,photonics2018}, optical communications~\cite{Optical_comm_2016} and image processing~\cite{SHE_image_proc2019}.

A similar spin-orbit coupling mechanism is expected to influence the propagation of wave packets in gravitational fields~\cite{GSHE_review,Oanceathesis}. This leads to gravitational spin Hall effects, which have been predicted for electromagnetic~\cite{GSHE2020,Harte_2022,Frolov2020,SHE_QM1}, linearized gravitational~\cite{GSHE_GW,SHE_GW} and Dirac~\cite{GSHE_Dirac,SHE_Dirac,rudiger,audretsch} fields propagating on curved spacetime backgrounds. In this general relativistic context, wave packets carrying intrinsic angular momentum follow frequency- and polarization-dependent trajectories when propagating through inhomogeneous gravitational fields. While spin Hall effects are generally observed in a wide range of physical systems that we can manipulate in our ground-based laboratories, these general relativistic predictions provide a novel avenue for probing spin-orbit interactions and spin Hall effects in the sky. As a comparison of spin Hall effects across these different areas of physics, we can say that black holes in spacetime play a similar role to the impurities of a semiconductor or to the inhomogeneities of an optical material. 

In this paper, we give a brief introduction to the gravitational spin Hall effect. We present the main steps in the derivation of the effect, discuss the analogies with optics, and give a brief overview of the potential astrophysical applications. Given the similarities between the gravitational spin Hall effects for different massless fields (electromagnetic, linearized gravity, massless Dirac), here we shall focus on the electromagnetic case, following the results presented in Refs.~\cite{GSHE2020,Harte_2022}. We start in Sec.~\ref{sec:SHElab} by reviewing the main features of spin Hall effects known from condensed matter physics and optics. A brief derivation of the spin Hall effect of light in inhomogeneous optical media is given here. This will help us introduce necessary concepts, such as the Berry phase, and will illustrate the strong analogies between the optical and gravitational case. In Sec.~\ref{sec:SHE_gravity}, we discuss the gravitational spin Hall effect of light. The effect is motivated by the well-known analogies between Maxwell's equations in curved spacetime and Maxwell's equations in optical media. Then, we present the main steps of the derivation of the effect, closely following the same method as in the optical case. Finally, we discuss some of the properties of the gravitational spin Hall equations and show how they reproduce several known results. In Sec.~\ref{sec:applications}, we discuss the main astrophysical applications of the gravitational spin Hall effect and the prospects for experimental observability. This discussion covers the strong-field lensing of gravitational waves and black hole shadows. Finally, we present our conclusions in Sec.~\ref{sec:conclusions}.

\section{Spin Hall effects in the lab} \label{sec:SHElab}

In this section, we briefly review the spin Hall effects observed in condensed matter physics and in optics. Given the strong similarities between the propagation of light in optical media and in curved spacetime, we will mainly focus on the spin Hall effect of light in an inhomogeneous medium. The main goal here is to present the main steps in the derivation of the spin Hall effect of light. This will help us introduce useful concepts, such as the Berry phase and the frequency- and polarization-dependent ray equations describing the propagation of light in inhomogeneous optical media beyond the geometrical optics approximation. We shall also discuss other relevant and related effects, such as the relativistic Hall effect and the geometric spin Hall effect of light.

We start in Sec.~\ref{sec:SHE_electrons} by briefly mentioning the spin Hall effect of electrons. We do not develop this topic in much detail, since our main interest throughout this paper is mainly towards spin Hall effects for massless fields. In Sec.~\ref{sec:SHEL}, we introduce the spin Hall effect of light. Given the striking similarities between electromagnetic waves propagating on curved spacetimes and electromagnetic waves inside some optical medium, this will serve as a foundation and motivation for the study of gravitational spin Hall effects. Finally, we discuss two frame-dependent effects for electromagnetic wave packets propagating in flat and empty spacetime: the relativistic Hall effect is discussed in Sec.~\ref{sec:wigner}, and the geometric spin Hall effect is presented in Sec.~\ref{sec:geometricSHE}.

\subsection{Spin Hall effect of electrons}
\label{sec:SHE_electrons}

In condensed matter physics, the spin Hall effect can be observed for electrons traveling in certain materials that exhibit spin-orbit coupling. This effect was first predicted by Dyakonov and Perel in 1971~\cite{originalSHE1,originalSHE2}, and describes the appearance of a spin current transverse to the electric charge current propagating in semiconductors. The effect was first observed by Bakun et al. in 1984~\cite{originalSHE3} as the inverse spin Hall effect, and in 2004 the direct spin Hall effect was observed in semiconductors~\cite{originalSHE4}. The source of this effect is the relativistic spin-orbit coupling between the electron spin and its center-of-mass motion inside a potential. Detailed reviews of the spin Hall effect of electrons can be found in Refs.~\cite{SHE_review, SHE_review1}.

\subsection{Spin Hall effect of light}
\label{sec:SHEL}

In optics, the spin Hall effect of light is generally present for polarized light propagating in a medium with an inhomogeneous refractive index or at the interface between two optical media, where the refractive index has a sharp jump. In this case, the polarization represents the spin internal degree of freedom. Light rays of opposite circular polarization are deflected in opposite directions, perpendicular to their direction of propagation and to the gradient of the refractive index~\cite{SOI_review}. Similarly to the case of electrons described above, inverse spin Hall effects of light can also occur in certain optical systems \cite{10.1038/ncomms6327, nayak2022momentum, bent_fibers}. In this case, the polarization of the electromagnetic wave evolves in a nontrivial way, controlled by the frequency and the trajectory of the beam.

The propagation of electromagnetic waves through various optical media is generally described by Maxwell's equations. While solving Maxwell's equations is important for the description of wave effects such as diffraction and interference, the vast majority of physical phenomena can also be described by applying certain approximations. For example, the propagation of light in optical media is often described within the geometrical optics approximation~\cite[Ch. III]{born1980}. In this setting, one assumes light of sufficiently high frequency, and Maxwell's equations, which are partial differential equations, are approximated by a set of ordinary differential equations, in the form of transport equations along rays. 

As a concrete example, consider the propagation of electromagnetic waves in a dielectric medium with an inhomogeneous refractive index $n(\mathbf{x})$ (for simplicity, we assume that the magnetic permeability is $\mu = 1$, and then the refractive index and the permittivity $\varepsilon$ are directly related as $n^2 = \varepsilon$). In this case, the wave equation for the electric field is~\cite[Sec. 1.2]{born1980}
\begin{equation}
    \nabla^2 \mathbf{E} - n^2 \frac{\partial^2 \mathbf{E}}{\partial t^2} + \nabla \left( \mathbf{E} \cdot \nabla \ln n^2 \right) = 0.
\end{equation}
The geometrical optics treatment starts with the following ansatz for the electric field:
\begin{equation}
    \mathbf{E} (\mathbf{x}, t) = \mathbf{E_0}(\mathbf{x}) e^{-i \omega t},
\end{equation}
where
\begin{equation}
    \mathbf{E_0}(\mathbf{x}) =  e^{i S(\mathbf{x})/\epsilon} \sum_{i=0} \epsilon^i \mathbf{e_i}(\mathbf{x}).
\end{equation}
In the above equations, $\omega$ is the frequency and $\epsilon$ is the wavelength. The main assumption of the geometrical optics approximation is that the wavelength $\epsilon$ is much smaller than the typical length scale of variation of the refractive index $n(\mathbf{x})$. We consider $\epsilon$ as a small expansion parameter and insert the above ansatz into the wave equation. At the lowest order in $\epsilon$, we obtain the eikonal equation (also called the geometrical optics dispersion relation)
\begin{equation} \label{eq:disp_go_maxwell}
    (\nabla S)^2 = n^2.
\end{equation}
This is a Hamilton-Jacobi equation for the phase function $S(\mathbf{x})$, which can be solved using the method of characteristics~\cite[Sec. 46]{Arnold_book}. In this case, the phase function $S(\mathbf{x})$ is determined by solving a set of ray equations, which are determined by the Hamiltonian function
\begin{equation}
    H(\mathbf{x}, \mathbf{p}) = \alpha \left[|\mathbf{p}|^2 - n^2(\mathbf{x}) \right] = 0.
\end{equation}
It is convenient to pick the constant $\alpha = |\mathbf{p} + n|^{-1}$ as in Ref.~\cite{Bliokh2009}, in which case $H = |\mathbf{p}| - n = 0$, and Hamilton's equations are 
\begin{equation} \label{eq:ray_go_maxwell}
    \dot{\mathbf{x}} = \frac{\mathbf{p}}{|\mathbf{p}|}, \qquad \dot{\mathbf{p}} = \nabla n(\mathbf{x}).
\end{equation}
These are the ray equations of geometrical optics in a medium with refractive index $n$. For the vast majority of physical applications, these equations provide a satisfactory model for the propagation of light in optical media. At this level of approximation, there are no polarization-dependent effects.

At the next order in the geometrical optics approximation, we obtain a transport equation for the amplitude vector $\mathbf{e_0}$~\cite[Sec. 3.1.3]{born1980}:
\begin{equation}
    \mathbf{\dot{e}_0} + \frac{1}{2} (\nabla^2 S) \mathbf{e_0} + (\mathbf{e_0} \cdot \nabla \ln n) \nabla S = 0,
\end{equation}
where the dot represents the derivative along the ray with tangent vector $\nabla S$. It is convenient to write the amplitude vector as $\mathbf{e_0} = \sqrt{I} \mathbf{u}$, where $I = \mathbf{\bar{e}_0} \cdot \mathbf{e_0}$ is the intensity, and $\mathbf{u} = \frac{\mathbf{e_0}}{\mathbf{\bar{e}_0} \cdot \mathbf{e_0}}$ is a unit polarization vector. Note that $e_0 \cdot \nabla S = 0$, which follows from Maxwell's equation $\nabla \cdot \mathbf{E} = 0$. Then, the above transport equation can be split into a transport equation for the intensity $I$ and a transport equation for the polarization vector $\mathbf{u}$. For the intensity, we have the following:
\begin{equation}
    \dot{I} = - I \nabla^2 S,
\end{equation}
This equation can be integrated along the light rays with tangent vector $\nabla S$ as
\begin{equation}
    I(\tau) = I(0) e^{-\int_0^\tau \nabla^2 S d\tau'}.
\end{equation}
The transport equation for the polarization vector $\mathbf{u}$ is
\begin{equation} \label{eq:transp_maxwell}
    \mathbf{\dot{u}} = - (\mathbf{u} \cdot \nabla \ln n) \nabla S.
\end{equation}
The dynamics of the polarization vector can be better understood in terms of a Berry phase and an associated Berry connection~\cite{Bliokh2008,Bliokh2009,SOI_review}. This can be done by introducing a frame of vectors adapted to the direction of propagation of light rays, $\mathbf{p} = \nabla S$. We consider a frame $(\mathbf{p}, \mathbf{v}, \mathbf{w})$ of $3$-vectors with real components, satisfying $\mathbf{p} \cdot \mathbf{v} = \mathbf{p} \cdot \mathbf{w} = \mathbf{v} \cdot \mathbf{w} = 0$ and $\mathbf{v} \cdot \mathbf{v} = \mathbf{w} \cdot \mathbf{w} = 1$. For example, this can be a Frenet-Serret frame, as considered in Ref.~\cite{Bliokh2008,Bliokh2009}, but other choices can also be made. From a physical point of view, one can think of the real vectors $\mathbf{v}$ and $\mathbf{w}$ as representing a linear polarization basis for the electric field. However, the dynamics of the polarization vector has a simplified form if we consider a circular polarization basis instead. Thus, we consider the frame $(\mathbf{p}, \mathbf{m}, \mathbf{\bar{m}})$ = $(\mathbf{p}, \frac{\mathbf{v} + i \mathbf{w}}{\sqrt{2}}, \frac{\mathbf{v} - i \mathbf{w}}{\sqrt{2}})$. Now, the polarization vector $\mathbf{u}$ can be expanded as
\begin{equation}
    \mathbf{u} = z_1 \mathbf{m} + z_2 \mathbf{\bar{m}},
\end{equation}
where $z_1$ and $z_2$ are two complex scalar functions. Inserting this expansion into the transport equation \eqref{eq:transp_maxwell}, we obtain
\begin{equation} \label{eq:transp_z_maxwell}
    \dot{z} = i \mathbf{B} \sigma_3 z,
\end{equation}
where $\sigma_3$ is the third Pauli matrix, $z$ is the unit two-dimensional complex vector
\begin{equation} \label{eq:z_def}
    z = \begin{pmatrix} z_1 \\ z_2 \end{pmatrix},
\end{equation}
and $\mathbf{B} = i \mathbf{\dot{\bar{m}}} \cdot \mathbf{m}$ represents a Berry connection. This describes the dynamics of the polarization vector $\mathbf{u}$ with respect to the frame $(\mathbf{p}, \mathbf{m}, \mathbf{\bar{m}})$. Note that the term on the right-hand side of Eq. \eqref{eq:transp_maxwell} does not contribute to the transport equation for $z$. However, we still see the effect of the inhomogeneous refractive index $n$ on the dynamics of $z$ -- changes in $n$ will result in changes in $\mathbf{p}$ and, consequently, in rotations of the adapted frame $(\mathbf{p}, \mathbf{m}, \mathbf{\bar{m}})$. Thus, the Berry connection $\mathbf{B}$ depends indirectly on the variations of $\mathbf{p}$ and the variations in the refractive index $n$.

The transport equation \eqref{eq:transp_z_maxwell} can be integrated as
\begin{equation}
    z(\tau) = \begin{pmatrix} e^{i \gamma(\tau)} & 0 \\ 0 & e^{- i \gamma(\tau)} \end{pmatrix} z(0),
\end{equation}
where $\gamma$ represents the Berry phase and is given by
\begin{equation}
    \gamma(\tau) = \int_0^\tau \mathbf{B} d\tau'.
\end{equation}
To summarize, from the lowest-order terms in the geometrical optics expansion we obtained the dispersion relation \eqref{eq:disp_go_maxwell} and the ray equations \eqref{eq:ray_go_maxwell}, while at the next order we obtained the transport equations along the geometrical optics rays for the intensity $I$ and for the polarization vector $\mathbf{u}$. The dynamics of the polarization vector can be studied in an adapted frame $(\mathbf{p}, \mathbf{m}, \mathbf{\bar{m}})$, which leads to a Berry connection and a Berry phase. While the Berry phase and the dynamics of the polarization vector in general depend on the geometrical optics rays \eqref{eq:ray_go_maxwell}, there is no backreaction from the state of polarization of the field onto the ray equations. In other words, spin-orbit couplings are not taken into account. 

To properly account for spin-orbit couplings, we follow Ref.~\cite{Bliokh2004} and we notice that for circularly polarized waves the electric field takes the form
\begin{equation}
    \mathbf{E_0} = \sqrt{I} \mathbf{m} e^{i(S+ \epsilon \gamma)/\epsilon} \quad \text{or} \quad  \mathbf{E_0} = \sqrt{I} \mathbf{\bar{m}} e^{i(S-\epsilon \gamma)/\epsilon}.
\end{equation}
Therefore, the total phase is $\tilde{S} = S + \epsilon s \gamma$, where $s = \pm 1$, depending on the state of circular polarization. Following Ref.~\cite{Bliokh2004}, spin-orbit couplings can be taken into account by treating the eikonal phase $S$ and the Berry phase $\gamma$ on an equal footing, and ray equations with polarization-dependent corrections can be obtained by deriving a modified dispersion relation for the total phase function $\tilde{S}$. This is equivalent to defining polarization-dependent effective refractive indices for each circular polarization state. Following~\cite{Bliokh2008,Bliokh2009} (an alternative derivation can also be found in Ref.~\cite{Ruiz2015}), we arrive at the modified ray equations that describe the spin Hall effect of light:
\begin{subequations} \label{eq:SHEL}
\begin{align}
    \dot{\mathbf{x}} &=  \frac{\mathbf{p}}{|\mathbf{p}|} + \epsilon s \frac{\mathbf{p} \times \dot{\mathbf{p}}}{|\mathbf{p}|^3}, \\
    \dot{\mathbf{p}} &= \nabla n(\mathbf{x}). 
\end{align}
\end{subequations}
Compared to the geometrical optic ray equations \eqref{eq:ray_go_maxwell}, these equations contain an additional term proportional to the wavelength $\epsilon$ and depending on the state of circular polarization of the light ray through $s = \pm 1$. Thus, in the limit of infinitely high frequency, we recover the results of geometrical optics. From a geometric point of view, the correction term can be related to the Berry curvature of the Berry connection~\cite{Bliokh2008,Bliokh2009,SOI_review}. The spin Hall effect of light, as described by the above equations, has been observed in several experiments~\cite{Hosten2008,Bliokh2008}. 

Our main goal in the following Sec.~\ref{sec:SHE_gravity}, motivated by the analogies between Maxwell's equations in inhomogeneous optical media and Maxwell's equations on curved spacetimes, will be to show how a similar effect arises for electromagnetic wave propagating under the influence of gravity.

\subsection{Relativistic Hall effect, Wigner translations and the observer dependence}
\label{sec:wigner}

Polarization-dependent effects can be seen even for electromagnetic beams propagating in vacuum Minkowski spacetime. The relativistic Hall effect introduced in Ref.~\cite{Relativistic_Hall} and the Wigner translations discussed in Ref.~\cite{Stone2015} describe the same effect, namely the observer dependence of the energy centroid of wave packets or beams carrying intrinsic angular momentum. In fact, this effect can even be seen for rotating mechanical bodies such as the flywheel discussed in Ref.~\cite{Relativistic_Hall}.   

The effect can be easily understood by considering a circularly polarized electromagnetic Gaussian beam propagating in the $z$ direction in Minkowski spacetime. A standard Minkowski observer with $4$-velocity $t^\alpha = (1, 0 ,0 ,0)$ will measure the energy density $T^{\alpha \beta} t_\alpha t_\beta$, where $T^{\alpha \beta}$ is the energy-momentum tensor of the beam. Then, at $z = t = 0$ the centroid of the beam in the $x-y$ plane can be computed as
\begin{equation}
    x_{c} = \frac{\int \int x T^{\alpha \beta} t_\alpha t_\beta dx dy }{ \int \int T^{\alpha \beta} t_\alpha t_\beta dx dy}, \qquad y_{c} = \frac{\int \int y T^{\alpha \beta} t_\alpha t_\beta dx dy }{ \int \int T^{\alpha \beta} t_\alpha t_\beta dx dy},
\end{equation}
In this case, the observer will see the energy centroid at the origin $(0, 0)$ of the $x - y$ plane. However, other observers may not agree with this. For example, an observer boosted in the $x$ direction, with $4$-velocity $\tilde{t}^\alpha = \frac{1}{\sqrt{1-v^2}} (1, v ,0 ,0)$ and $v < c = 1$ will measure the energy density $T^{\alpha \beta} \tilde{t}_\alpha \tilde{t}_\beta$, and the energy centroid will now be shifted in the $y$-direction~\cite{Stone2015}:
\begin{equation}
    (\tilde{x}_c, \tilde{y}_c) = \left( 0, \epsilon s v + \mathcal{O}(\epsilon^2) \right),
\end{equation}
where $\epsilon$ is the wavelength and $s = \pm 1$, depending on the circular polarization state of the electromagnetic beam. Since the boost velocity $v$ is limited by the speed of light, the size of the shift is limited to one wavelength and the direction of the shift is controlled by the direction of the intrinsic angular momentum carried by the beam. A graphical representation of this effect can be found in Ref.~\cite[Fig. 2]{Stone2015}. 

So far, it is not clear how the relativistic Hall effect is connected to the spin Hall effect of light discussed in the previous section. This is because a standard Minkowski observer with $4$-velocity $t^\alpha = (1, 0 ,0 ,0)$ was secretly assumed when deriving the spin Hall equations \eqref{eq:SHEL}. This connection will be clarified in Sec.~\ref{sec:SHE_gravity} where we derive the covariant equations describing the gravitational spin Hall effect and where the observer plays an important role.

\subsection{Geometric spin Hall effect of light}
\label{sec:geometricSHE}

Another polarization-dependent effect observed for electromagnetic beams is the geometric spin Hall effect of light~\cite{Aiello2009,Korger2011,Korger2014}. This effect occurs when a polarized light beam is observed on a two-dimensional screen tilted with respect to the direction of propagation of the beam. The beam centroid will be shifted in a direction perpendicular to the direction of propagation of the beam, the size of the effect is proportional to the wavelength and to the tangent of the tilt angle between the beam and the screen, and the direction of the shift can be controlled by the state of polarization of the beam. The geometric spin Hall effect of light was observed experimentally in Ref.~\cite{Korger2014}. However, there is no known connection between this effect and the gravitational spin Hall effect, so we will not discuss this further.

\section{Spin Hall effects in general relativity} \label{sec:SHE_gravity}

In this section, we give an overview of the derivation and the main properties of the gravitational spin Hall effect. Spin-dependent effects on the propagation of wave packets and particles have been widely studied in general relativity using different methods, such as WKB-type approximations~\cite{GSHE2020,Frolov2020,GSHE_GW,Oanceathesis,Harte_2022,GSHE_Dirac,Frolov, Frolov2, covariantSpinoptics, Harte2018, spinorSpinoptics, spinorSpinoptics2, shoom2020, audretsch,rudiger}, the Mathisson-Papapetrou equations for spinning objects~\cite{souriau1974modele, saturnini1976modele, Duval,Duval2018,Duval2019,Harte_2022}, approximation methods inspired from quantum mechanics~\cite{SHE_QM1,SHE_Dirac,SHE_GW} and others~\cite{PhysRevD.105.096019,PhysRevD.105.104008} (see also Refs.~\cite{FaradayRotation1,FaradayRotation2,FaradayRotation3,FaradayRotation4,FaradayRotation5,FaradayRotation6,Li:2022izh}). Furthermore, gravitational spin Hall effects are predicted for different fields propagating in curved spacetime, such as electromagnetic~\cite{GSHE2020,Frolov2020,Harte_2022,SHE_QM1}, linearized gravitational~\cite{GSHE_GW,SHE_GW} and massive and massless Dirac fields~\cite{GSHE_Dirac,rudiger,audretsch} (spin Hall effects have also been predicted for more exotic particles such as massless particles with anyonic spin~\cite{marsot2022,gray2022,marsot2022hall,bicak2023}). 

Based on astrophysical relevance, our main focus here will be on gravitational spin Hall effects for massless fields. Furthermore, given the similarities between the derivations and the resulting spin Hall equations of motion for different massless fields, we shall restrict our attention to the electromagnetic case discussed in Refs.~\cite{GSHE2020,Harte_2022}.

We start in Sec.~\ref{sec:analogue} with the analogy between the description of electromagnetic waves in curved spacetimes and in optical materials. Based on the known results from optics, this should serve as a strong motivation towards gravitational spin Hall effects. Then, in Sec.~\ref{sec:GSHE_light} we present the main steps of the derivation of the gravitational spin Hall effect of light, based on a WKB approach. This results in a set of modified ray equations, similar to the ones derived in optics, which contain polarization and frequency-dependent correction terms to the geodesic equations of geometrical optics. In Sec.~\ref{sec:MPD}, we show that these modified ray equations can be understood as a particular case of the Mathisson-Papapetrou equations for spinning particles. In Sec.~\ref{sec:optical_metric}, we again use the analogy between optical materials and curved spacetime, this time in the form of the optical metric. In this way, the spin Hall effect of light in optics can be recovered from the gravitational spin Hall effect of light.

\subsection{Analogies between curved spacetimes and optical materials} \label{sec:analogue}

Analogies between the propagation of light in curved spacetimes and in optical materials had already been recognized from the early days of general relativity. In his book published in 1920~\cite{Eddington}, Eddington mentioned the possibility of describing gravitational light bending around the Sun by considering space to be filled with a refractive medium. This analogy was later developed by Gordon~\cite{Gordon}, Skrotskii~\cite{skrotskii} and Plebanski~\cite{Plebansky-Maxwell} (see also Refs.~\cite{deFelice,cartographic_analog}). 

Based on Plebanski's approach given in Ref.~\cite{Plebansky-Maxwell}, we can see that Maxwell's equations in a vacuum spacetime described by the metric tensor $g_{\mu \nu}$ are equivalent to the flat spacetime Maxwell's equations inside an optical medium with perfect impedance matching. The analogue material is described by the constitutive equations
\begin{subequations}
\begin{align}
    \mathbf{D} &= \mathbf{\epsilon} \mathbf{E} + \mathbf{\gamma} \mathbf{H}, \\
    \mathbf{B} &= \mathbf{\mu} \mathbf{H} - \mathbf{\gamma} \mathbf{E},
\end{align}
\end{subequations}
where the properties of the material are described by tensorial quantities, related to the metric components as
\begin{equation}
\epsilon^{i j} = \mu^{i j} = -\sqrt{-\det g} \frac{g^{i j}}{g_{0 0}}, 
 \qquad  \qquad
\gamma^{i j} =- \varepsilon^{i j k} \frac{g_{0 k}}{g_{0 0}}.
\end{equation}
This can be seen as an example of analogue gravity~\cite{Analogue_gravity}, where certain properties of a curved spacetime can be reproduced by other physical systems. Note that, while at least in principle we can define an analogue optical medium for any Lorentzian spacetime metric tensor $g_{\mu \nu}$, there are many optical materials that do not have a Lorentzian spacetime analogue (for example, the impedance matching condition $\epsilon^{i j} = \mu^{i j}$ is very restrictive, and the vast majority of materials commonly used in optics do not satisfy this property).

The analogy presented above, together with the theoretical prediction and experimental observation of the spin Hall effect of light in inhomogeneous optical media, strongly suggests that a similar effect should also affect the propagation of electromagnetic waves in curved spacetime. 

\subsection{Gravitational spin Hall effect -- a WKB approach} \label{sec:GSHE_light}

In this section, we present the main steps of the derivation of the gravitational spin Hall effect. A general approach for the description of this effect has been given for electromagnetic waves in Ref.~\cite{GSHE2020}, and later the same method has been adapted for linearized gravitational waves in Ref.~\cite{GSHE_GW} and for Dirac fields in Ref.~\cite{GSHE_Dirac}. Here, we mainly focus on the derivation of the gravitational spin Hall effect of light, as given in Refs.~\cite{GSHE2020, Harte_2022}. The derivation of the effect in the case of linearized gravitational waves~\cite{GSHE_GW} or Dirac fields~\cite{GSHE_Dirac} can be performed following similar steps.

We assume a fixed spacetime background and we write Maxwell's equations for the vector potential
\begin{equation}
    \left( \nabla^\beta \nabla_\alpha - \delta^\beta_\alpha \nabla^\gamma \nabla_\gamma \right) A_\beta = 0.
\end{equation}
The gauge is fixed by imposing the Lorenz gauge condition 
\begin{equation}
    \nabla_\alpha A^\alpha = 0.
\end{equation}
We aim to describe the propagation of electromagnetic waves of large but finite frequencies. Thus, we assume that the wavelength of the electromagnetic wave is much smaller than the typical length scale of variation of the gravitational field, and that the vector potential admits a WKB expansion of the form
\begin{equation}
    A_\alpha =  \mathrm{Re} \left[ \left( {A_0}_\alpha + \epsilon {A_1}_\alpha + \mathcal{O}(\epsilon^2) \right) e^{i S/\epsilon} \right], 
    \label{WKB0}
\end{equation}
where $\epsilon$ is a small expansion parameter related to the wavelength, $S$ is a real phase function and ${A_i}_\alpha$ are complex amplitudes. It is convenient to define a wave vector $k_\alpha = \nabla_\alpha S$, and then a timelike observer with $4$-velocity $t^\alpha$ will measure a wave frequency
\begin{equation}
    \omega = - \frac{k_\alpha t^\alpha}{\epsilon}.
\end{equation}
Next, following similar steps as in Sec.~\ref{sec:SHEL}, we insert the WKB ansatz into Maxwell's equations, as well as the Lorenz gauge condition. At the lowest order in $\epsilon$, we obtain
\begin{subequations}
\begin{align}
    k_\alpha k^\alpha &= 0, \label{eq:GO_disp}\\
    \qquad k^\alpha {A_0}_\alpha &= 0. \label{eq:orthogonality}
\end{align}
\end{subequations}
The first equation represents the geometrical optics dispersion relation, which is a Hamilton-Jacobi equation for the phase function $S$. This can be solved by using the method of characteristics~\cite[Sec. 46]{Arnold_book}, in which case we define a Hamiltonian function 
\begin{equation}
    H(x, p) = \frac{1}{2} g^{\alpha \beta} p_\alpha p_\beta = 0,
\end{equation}
and Hamilton's equations are the geodesic equations of the background spacetime:
\begin{subequations} \label{eq:geodesics}
\begin{align}
    \dot{x}^\mu &= \frac{\partial H}{\partial p_\mu} = p^\mu, \\
    \dot{p}_\mu &= -\frac{\partial H}{\partial x^\mu} =\Gamma^\alpha_{\beta \mu} p_\alpha p^\beta.
\end{align}
\end{subequations}
This represents the well-known geometrical optics result that light rays follow null geodesics~\cite{MTW}. Alternatively, the geodesic equations can also be derived directly from the geometrical optics dispersion relation \eqref{eq:GO_disp} by differentiation:
\begin{equation}
    0 = \nabla_\mu \left(\frac{1}{2} k_\alpha k^\alpha \right) = k^\alpha \nabla_\mu k_\alpha = k^\alpha \nabla_\alpha k_\mu.
\end{equation}
The second equation, \eqref{eq:orthogonality}, arises from the Lorenz gauge condition and requires that the wave vector $k_\alpha$ and the amplitude vector ${A_0}_\alpha$ to be orthogonal. This means that the amplitude vector ${A_0}_\alpha$ must belong to the space orthogonal to the null vector $k_\mu$. To emphasize this result, we define an adapted null tetrad $\{ k_\alpha, t_\alpha, m_\alpha, \bar{m}_\alpha \}$, where $k_\alpha$ is the null wave vector, $t_\alpha$ is a real timelike vectors, $m_\alpha$ and $\bar{m}_\alpha$ are complex null vectors, and the only nonzero contractions are $t_\alpha t^\alpha = -1$, $k_\alpha t^\alpha = -\epsilon \omega$ and $m_\alpha \bar{m}^\alpha = 1$. Thus, the amplitude vector can be expanded as
\begin{equation} \label{eq:field_expansion}
    {A_0}_\alpha = \sqrt{I} a_\alpha = \sqrt{I} \left( z_1 m_\alpha + z_2 \bar{m}_\alpha + z_3 k_\alpha \right),
\end{equation}
where $I = {\bar{A}_0}^{\alpha} {A_0}_\alpha$ is a real intensity, $a_\alpha$ is a unit-complex polarization vector and $z_i$ are complex scalar functions. The values of $z_1$ and $z_2$ encode the state of polarization of the field, while the term proportional to $z_3$ is a residual gauge term that is not fixed by the Lorenz gauge. The complex null vectors $m_\alpha$ and $\bar{m}_\alpha$ represent a circular polarization basis, and we have circularly polarized electromagnetic waves when $z_1 = 0$ or $z_2 = 0$.

At the next-to-leading order of the WKB expansion, we obtain the following transport equation for the complex amplitude vector:
\begin{equation}
    k^\mu \nabla_\mu {A_0}_\alpha + \frac{1}{2} {A_0}_\alpha \nabla_\mu k^\mu = 0.
\end{equation}
Using Eq. \eqref{eq:field_expansion}, we can split this into a transport equation for the intensity $I$ and a transport equation for the polarization vector $a_\alpha$:
\begin{subequations}
\begin{align}
    \nabla_\mu \left( I k^\mu \right) &= 0, \\
    k^\mu \nabla_\mu a_\alpha &= 0.
\end{align}
\end{subequations}
The transport equation for the intensity can be integrated along light rays $x^\mu(\tau)$ with the tangent vector $\dot{x}^\mu = k^\mu$ as 
\begin{equation}
    I(\tau) = I(0) e^{- \int_0^\tau \nabla_\mu k^\mu d\tau'}.
\end{equation}
More interesting to us is the parallel transport equation for the polarization vector $a_\mu$. Similarly to the approach used in optics, it is convenient to expand the polarization vector in an adapted null tetrad $\{ k_\alpha, t_\alpha, m_\alpha, \bar{m}_\alpha \}$ as in Eq. \eqref{eq:field_expansion}. Then the parallel transport equation can be rewritten as transport equations for the scalar functions $z_1$ and $z_2$ (there is also a corresponding transport equation for $z_3$, but we ignore this, as it only describes the dynamics of a gauge term):
\begin{equation} \label{eq:zdot}
    \dot{z} = i k^\alpha B_\alpha \sigma_3 z ,
\end{equation}
where $z$ has the same form as in Eq. \eqref{eq:z_def}, and the Berry connection $B_\alpha$ is defined as
\begin{equation}
\begin{split}
    B_\alpha (x, k) &= i \bar{m}^\beta (x, k) \nabla_\alpha\left[ m_\beta(x,k) \right] \\
    &= i \bar{m}^\beta  \left( \nabla_\alpha +  k_\gamma \Gamma^\gamma_{\alpha \lambda} \frac{\partial}{\partial k_\lambda} \right) m_\beta.
\end{split}
\end{equation}
In the above equation, we used the fact that $m_\alpha$ and $\bar{m}_\alpha$ are functions of the coordinates $x^\mu$, as well as the wave vector $k_\mu(x)$. This is due to the orthogonality relations that define the adapted null tetrad $\{ k_\alpha, t_\alpha, m_\alpha, \bar{m}_\alpha \}$. Any change in $k_\alpha$ will result in a change in the orthogonal vectors $m_\alpha$ and $\bar{m}_\alpha$. Thus, the second line of the above equation simply follows from applying the chain rule, and the operator in the brackets represents a covariant horizontal derivative on the cotangent bundle. The transport equation \eqref{eq:zdot} can be integrated, and we obtain
\begin{equation}
    z(\tau) = \begin{pmatrix} e^{i \gamma(\tau)} & 0 \\ 0 & e^{- i \gamma(\tau)} \end{pmatrix} z(0),
\end{equation}
where $\gamma$ is the Berry phase, defined as
\begin{equation}
    \gamma(\tau) = \int_{0}^\tau d\tau' k^\alpha B_\alpha.
\end{equation}
Thus, the polarization dynamics for electromagnetic waves in curved spacetimes or in optical media can be described in very similar terms. In both cases, the Berry connection and the Berry phase arise by expanding the polarization vector in terms of a frame adapted to the wave vector. A similar result also holds for linearized gravitational waves, the only difference being that the Berry phase will include an additional factor of $2$~\cite{GSHE_GW}. This is because gravitational waves are represented by a massless spin-$2$ field, whereas electromagnetic waves have spin-$1$. A similar transport equation defined in terms of a Berry connection was also obtained in the case of Dirac fields~\cite{GSHE_Dirac}.

At this stage, spin-orbit interactions are not taken into account. The dynamics of light rays is described by the null geodesic equations, and the dynamics of the polarization along the null geodesics is described by the Berry phase. While the polarization is affected by the followed light rays, there is no backreaction from the state of polarization of the field onto the ray dynamics. The first step towards including spin-orbit interactions into the dynamics is to note that the Berry phase can be viewed as a higher-order correction to the total phase of the WKB field. More concretely, for circularly polarized electromagnetic waves, the WKB field takes the form (here, we ignore the residual gauge terms proportional to $k_\alpha$ in the polarization vector)
\begin{equation}
    A_\alpha = \mathrm{Re} \left[ \left(\sqrt{I} m_\alpha + \mathcal{O}(\epsilon)\right) e^{i (S+\epsilon \gamma)/\epsilon} \right] \quad \text{or} \quad A_\alpha =  \mathrm{Re} \left[ \left(\sqrt{I} \bar{m}_\alpha + \mathcal{O}(\epsilon)\right) e^{i (S-\epsilon \gamma)/\epsilon}\right].
\end{equation}
Thus, the total phase is $\tilde{S} = S + \epsilon s \gamma$, where $s = \pm 1$ depends on the state of circular polarization of the field. The geometrical optics ray equations \eqref{eq:geodesics} were derived considering the Hamilton-Jacobi equation for the phase function $S$, defined by the geometrical optics dispersion relation \eqref{eq:GO_disp}. Following Refs.~\cite{GSHE2020,Bliokh2004}, spin-orbit interactions can be taken into account by including the Berry phase $\gamma$ as a higher-order correction to the geometrical optics phase function $S$. Thus, we can derive an effective dispersion relation for the total phase function $\tilde{S}$. Using the results obtained above, we get
\begin{equation}
     \frac{1}{2} (\nabla_\mu \tilde{S}) (\nabla^\mu \tilde{S}) - \epsilon s (\nabla^\mu \tilde{S}) B_\mu = \mathcal{O}(\epsilon^2).
\end{equation}
The above effective dispersion relation contains a higher-order term proportional to the Berry connection and depending on the state of circular polarization of the field through $s = \pm 1$. We can treat this as a Hamilton-Jacobi equation for the total phase function $\tilde{S}$, and applying the method of characteristics leads us to the effective Hamiltonian function 
\begin{equation}
    H (x, p) = \frac{1}{2} g^{\alpha \beta} p_\alpha p_\beta - \epsilon s p^\mu B_\mu(x, p) + \mathcal{O}(\epsilon^2) = 0. 
\end{equation}
In principle, we could just use the above Hamiltonian, together with the standard symplectic $2$-form $\Omega = dx^\mu \wedge dp_\mu$, to describe the gravitational spin Hall effect of light, and the corresponding Hamilton's equations would consist of effective ray equations with polarization-dependent corrections to the null geodesics. However, the above Hamiltonian explicitly depends on the Berry connection, which is a gauge-dependent term. Here, the gauge dependence does not refer to the gauge of the electromagnetic vector potential $A_\alpha$, but rather to the fact that the Berry connection depends on the choice of complex null vectors $m_\alpha$ and $\bar{m}_\alpha$. Given a null wave vector $k_\alpha$, and additionally picking a timelike vector $t_\alpha$ (or, alternatively, another null vector $n_\alpha$), the complex null vectors $m_\alpha$ and $\bar{m}_\alpha$ are fixed only up to a spin rotation $m_\alpha \mapsto e^{i \phi} m_\alpha$. The Berry connection, as well as the Berry phase, is not invariant under spin rotations, transforming as $B_\alpha \mapsto B_\alpha - \nabla_\alpha \phi$ and $\gamma(\tau) \mapsto \gamma(\tau) - \phi(\tau) + \phi(0)$. Thus, we would like to remove such gauge-dependent terms from the Hamiltonian, and to arrive at a gauge-invariant description of the gravitational spin Hall effect of light. Generally, this can be done by performing a coordinate transformation of the type introduced in Ref.~\cite{Littlejohn1991}. In our case, we perform the following coordinate transformation:
\begin{subequations}
\begin{align} \label{eq:coord}
    x^\mu &\mapsto x^\mu - i \epsilon s \bar{m}^{\alpha} \frac{\partial}{\partial p_\mu} {m}_\alpha, \\
    p_\mu &\mapsto p_\mu + i \epsilon s \bar{m}^{\alpha} \nabla_\mu {m}_\alpha.
\end{align}
\end{subequations}
In the new coordinates, the Hamiltonian function can be written as
\begin{equation}
    H (x, p) = \frac{1}{2} g^{\alpha \beta} p_\alpha p_\beta + \mathcal{O}(\epsilon^2) = 0,
\end{equation}
and the symplectic $2$-form becomes
\begin{equation}
    \Omega = dx^\mu \wedge dp_\mu + F,
\end{equation}
where $F$ is the Berry curvature $2$-form associated with the Berry connection $B_\mu$, with components
\begin{subequations}
\begin{align}
    (F_{pp})^{\beta\alpha} &= 2 \epsilon s \mathrm{Im} \left( \frac{ \partial  m_\gamma }{ \partial p_{\alpha} } \frac{ \partial \bar{m}^\gamma }{ \partial p_{\beta} }  \right),
    \\
    (F_{xx})_{\beta\alpha} &= 2 \epsilon s \mathrm{Im} \left(\nabla_{\alpha} m_\gamma \nabla_{\beta} \bar{m}^{\gamma} +  m_\gamma \nabla_{[\alpha} \nabla_{\beta]} \bar{m}^\gamma \right),
    \\
    (F_{px})_{\alpha}{}^{\beta} &= - (F_{xp})^{\beta}{}_{\alpha} 
    = 2 \epsilon s \mathrm{Im} \left( \frac{ \partial m_\gamma }{ \partial p_{\beta}} \nabla_\alpha \bar{m}^\gamma \right).
\end{align}
\end{subequations}
Thus, we have successfully removed the gauge-dependent terms from the Hamiltonian, and now the gravitational spin Hall effect is described in terms of the gauge-invariant Berry curvature $F$, which appears as a correction term in the symplectic $2$-form $\Omega$. Furthermore, using the properties of the adapted null tetrad $\{ k_\alpha = p_\alpha, t_\alpha, m_\alpha, \bar{m}_\alpha \}$, the components of the Berry curvature can be rewritten as 
\begin{subequations} \label{eq:Berry_curvature}
\begin{align}
    \left({F_{p p}}\right)^{\beta \alpha} &= \frac{ S^{\alpha \beta} }{(p \cdot t)^2} , 
    \\
    \left({F_{xx}}\right)_{\beta \alpha} &=  \frac{ S^{\gamma \lambda }}{2} \bigg[ R_{\gamma  \lambda \alpha \beta} + \frac{2}{ (p \cdot t)^2}  p_\rho \Gamma^\rho_{\gamma [ \alpha } \bigg( \Gamma^\sigma_{\beta] \lambda} p_\sigma - 2 (p \cdot t)  \nabla_{\beta]} t_\lambda  \bigg) \bigg],
    \\
    \left({F_{x p}}\right)\indices{^\alpha_\beta} &= \frac{ S^{\alpha \gamma}  }{(p \cdot t)^2}  \left( p_\rho \Gamma^\rho_{\beta \gamma} -( p \cdot t ) \nabla_\beta t_\gamma \right).
\end{align}
\end{subequations}
All components of the Berry curvature are linear in the spin tensor $S^{\alpha \beta}$, defined as
\begin{equation} \label{eq:spin_tensor}
   S^{\alpha \beta} = 2 i \epsilon s \bar{m}^{[\alpha} m^{\beta]} = \epsilon s \frac{ \varepsilon^{\alpha \beta \mu \nu} p_\mu t_\nu }{p_\sigma t^\sigma}.
\end{equation}
Note that, up to the sign of $s = \pm 1$, the spin tensor is uniquely determined by $p_\alpha$ and $t_\alpha$. The spin tensor encodes the amount of angular momentum carried by the electromagnetic wave packet. Finally, we can use the above Hamiltonian and symplectic $2$-form to write the gauge-invariant equations of motion describing the gravitational spin Hall effect of light:
\begin{subequations} \label{eq:gshe_eq}
    \begin{align}
    \dot{x}^\alpha &= p^\alpha + \frac{1}{p_\sigma t^\sigma} S^{\alpha \beta} p^\mu \nabla_\mu t_\beta, \label{eq:xdot}
    \\
    \dot{x}^\mu \nabla_\mu p_\alpha  &= -  \frac{1}{2} R_{\alpha \beta \gamma \lambda}  p^\beta  S^{\gamma \lambda} . \label{eq:pdot}
    \end{align}
\end{subequations}
These equations are similar to the ones derived in optics in Eq. \eqref{eq:SHEL}, containing frequency and polarization-dependent correction terms to the geometrical optics null geodesics. The spin Hall effect vanishes in the limit of infinitely high frequency, and we recover geometrical optics. The physical meaning of these equations is that, in the intermediate regime of large but finite frequencies, the dynamics of circularly polarized electromagnetic wave packets deviates from the prediction of geometrical optics. The worldline $x^\mu(\tau)$ represents the energy centroid and $p_\mu(\tau)$ represents the average momentum of the wave packet, while the spin tensor $S^{\alpha \beta}$ describes the internal degree of freedom of the wave packet. Furthermore, note that $S^{\alpha \beta}$, as well as the equations of motion, depend on the external choice of a timelike vector field $t^\alpha$. This should come as no surprise, as the energy centroid of a wave packet is an observer-dependent quantity. The timelike vector field $t^\alpha$ should be viewed as the $4$-velocity of a family of observers that describe the motion of the wave packet. Thus, the equations of motion depending on $t^\alpha$ reflect the fact that different observers will assign different centroids for the same wave packet, as was also the case for the relativistic Hall effect and the Wigner translations discussed in Sec.~\ref{sec:wigner}. As a result of the covariant approach used here, the gravitational spin Hall equations of motion \eqref{eq:gshe_eq} automatically include these observer-dependent effects, providing a unified framework for spin Hall effects and relativistic Hall effects.

The gravitational spin Hall equations for linearized gravitational waves~\cite{GSHE_GW} and for massless Dirac fields~\cite{GSHE_Dirac} take the same form as Eq. \eqref{eq:gshe_eq}. The only difference is a constant numerical factor in the definition of the spin tensor, which represents the fact that these are fields of different spin and hence carry different amounts of angular momentum. For linearized gravitational waves we have $s = \pm 2$ and for massless Dirac fields we have $s = \pm \frac{1}{2}$. In the case of Dirac fields, the gravitational spin Hall equations can also be extended to include the effect of charge and an externally applied electromagnetic field~\cite{GSHE_Dirac}.

\begin{figure}
    \centering
    \includegraphics[width=0.9\columnwidth]{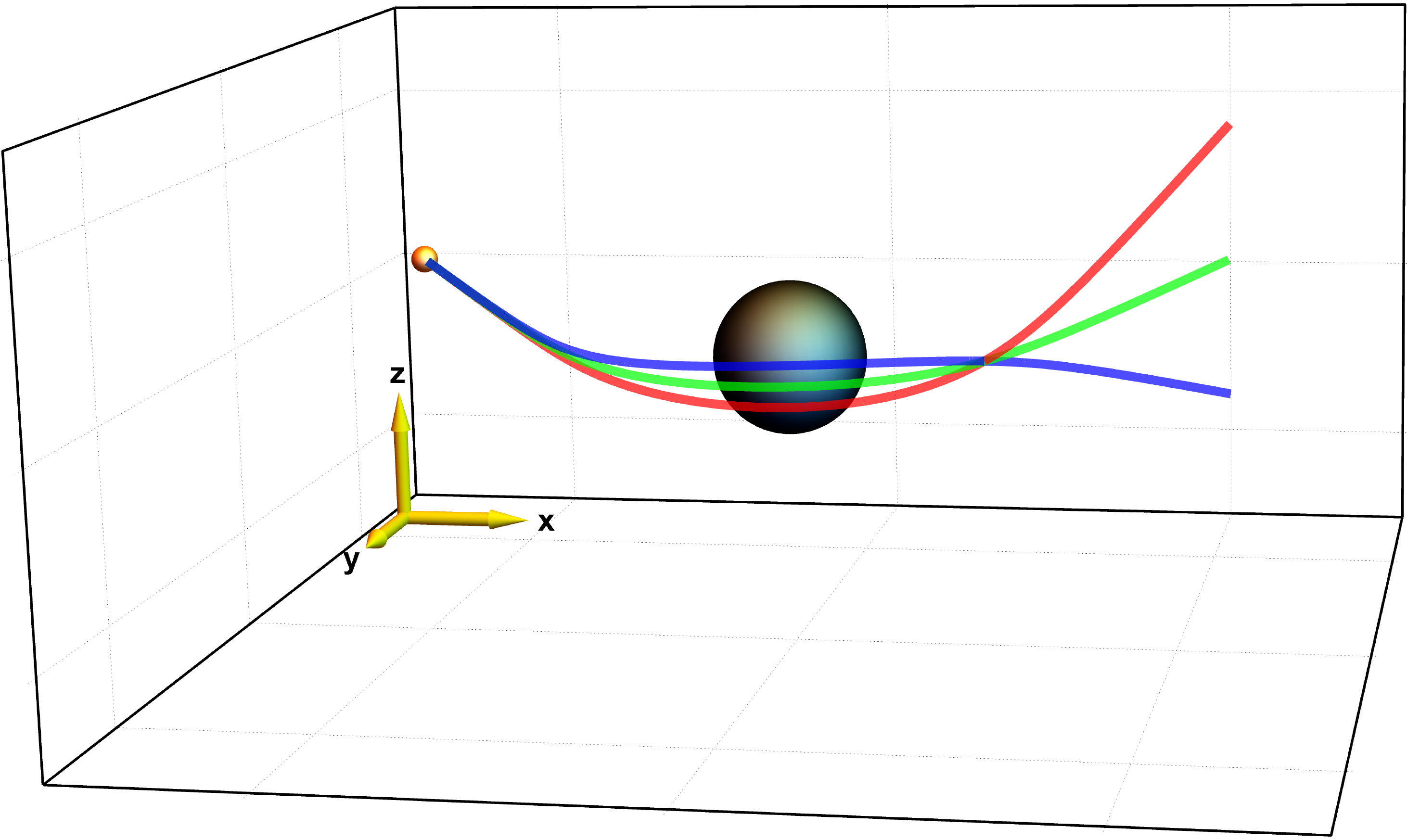}
    \caption{The gravitational spin Hall effect of light around a Schwarzschild black hole. A source of light, represented by the small yellow sphere, is considered in the equatorial plane of the black hole. Three light rays are emitted by the source with the same initial conditions $(x^\alpha(0), p_\alpha(0))$, but with different wavelengths $\epsilon$ and polarizations. The green line corresponds to $\epsilon = 0$ and represents a null geodesic. The red and the blue lines represent the trajectories of polarized light rays with $s = \pm 1$ and with the same finite wavelength $\epsilon$. The magnitude of the effect is exaggerated for visualisation purposes.}
    \label{fig:gshe_example}
\end{figure}

More intuition about the gravitational spin Hall effect can be gained by numerically integrating Eq. \eqref{eq:gshe_eq}. As a basic example, let us consider a Schwarzschild black hole, together with a light source close by. Then, we can emit several light rays with the same initial conditions $(x^\alpha(0), p_\alpha(0))$ for Eq. \eqref{eq:gshe_eq}, but with different wavelengths and polarizations. This example is presented in Fig. \ref{fig:gshe_example}. For $\epsilon = 0$, we obtain a null geodesic that will have a bent trajectory, but its motion remains confined to the equatorial plane. The gravitational spin Hall effect is obtained when looking at light rays of finite wavelength $\epsilon$. In this case, we obtain two different trajectories corresponding to light rays of opposite circular polarization ($s = \pm 1$). These modified trajectories are no longer confined to the equatorial plane, and, in addition to the ordinary light bending of null geodesics, the gravitational spin Hall effect induces an additional deflection angle in a direction orthogonal to the geodesic plane. Other trajectory examples can be found in Refs.~\cite{GSHE2020,Oanceathesis,GSHE_lensing}. Furthermore, a Mathematica notebook that numerically integrates Eq. \eqref{eq:gshe_eq} and plots the gravitational spin Hall trajectories can be found in Ref.~\cite[Appendix A.7]{Oanceathesis}.

\subsection{Comparison with the Mathisson-Papapetrou equations} \label{sec:MPD}

In this section, we show how the gravitational spin Hall equations of motion \eqref{eq:gshe_eq} are related to the well-known Mathisson-Papapetrou equations for spinning objects in general relativity. This correspondence, established in Ref.~\cite{Harte_2022}, enables us to use several known results for the Mathisson-Papapetrou equations and apply them for the study of the gravitational spin Hall effect.

In general relativity, the motion of compact spinning objects with conserved energy-momentum tensor can be described using the Mathisson-Papapetrou equations~\cite{Mathisson,Papapetrou,Dixon74,dixon2015new}
\begin{subequations} \label{eq:MP}
\begin{align}
    \dot{x}^\mu \nabla_\mu p_\alpha  &= - \frac{1}{2}  R_{\alpha \beta \gamma \lambda } \dot{x}^\beta S^{\gamma \lambda}, 
    \label{MPp}
    \\
    \dot{x}^\mu \nabla_\mu S^{\alpha \beta} &=  p^{\alpha} \dot{x}^{\beta} - p^{\beta} \dot{x}^{\alpha}.
    \label{MPS}
\end{align}
\end{subequations}
Since electromagnetic waves satisfy the conservation law for the energy-momentum tensor, and we are mainly interested in sufficiently localized wave packets, the Mathisson-Papapetrou equations should give a valid description for the dynamics of such objects. First, note that the Mathisson-Papapetrou equations are evolution equations for $p_\alpha$ and $S^{\alpha \beta}$ along a fixed worldline $x^\mu(\tau)$. However, there is no evolution equation to determine the worldline and the system of equations \eqref{eq:MP} is underdetermined. An evolution equation for the worldline can be obtained by imposing a spin supplementary condition~\cite{Costa2015}, which generally takes the form $S^{\alpha \beta} u_\beta = 0$ for some chosen vector $u_\beta$. 

It has been shown in Ref.~\cite{Harte_2022} that the gravitational spin Hall equations \eqref{eq:gshe_eq} are a special case of the Mathisson-Papapetrou equations. This follows by fixing a timelike vector field $t^\alpha$ and using the Corinaldesi-Papapetrou spin supplementary condition~\cite{CP_ssc, Costa2015}
\begin{equation} \label{eq:ssc}
    S^{\alpha \beta} t_\beta = 0.
\end{equation}
Furthermore, the worldline parameter $\tau$ has to be chosen such that $\dot{x}^\mu t_\mu = p^\mu t_\mu$, and we need to impose the initial conditions 
\begin{equation} \label{eq:initial_cond}
    S^{\alpha \beta} p_\beta = 0, \qquad S^{\alpha \beta} S_{\alpha \beta} = 2 (\epsilon s)^2.
\end{equation}
The evolution equation for the worldline follows from Eqs. \eqref{MPS} and \eqref{eq:ssc} as
\begin{equation}
\begin{split}
    0 &= \dot{x}^\mu \nabla_\mu \left( S^{\alpha \beta} t_\beta \right) \\
    &= p^\alpha \dot{x}^\beta t_\beta - p^\beta t_\beta \dot{x}^\alpha + S^{\alpha \beta} \dot{x}^\mu \nabla_\mu t_\beta.
\end{split}
\end{equation}
Using a choice of parametrization such that $\dot{x}^\mu t_\mu = p^\mu t_\mu$, we obtain
\begin{equation}
    \dot{x}^\alpha = p^\alpha + \frac{1}{p_\sigma t^\sigma} S^{\alpha \beta} \dot{x}^\mu \nabla_\mu t_\beta.
\end{equation}
Recall that the spin Hall equations \eqref{eq:gshe_eq} are correct only up to error terms of order $\epsilon^2$. Thus, we can use $\dot{x}^\mu = p^\mu + \mathcal{O}(\epsilon)$ and then the above equation is equivalent to Eq. \eqref{eq:xdot}. Similarly, the Mathisson-Papapetrou equation \eqref{MPp} for the momentum is also equivalent, up to error terms of order $\epsilon^2$ with the spin Hall equation \eqref{eq:pdot}. The spin supplementary condition \eqref{eq:ssc}, together with the initial conditions given in Eq. \eqref{eq:initial_cond}, fix the spin tensor to be of the form given in Eq. \eqref{eq:spin_tensor}. Then, it can be shown that this spin tensor satisfies the Mathisson-Papapetrou equation $\eqref{MPS}$.

The established correspondence between the Mathisson-Papapetrou equations and the gravitational spin Hall equations allows us to use several known results for the former and use them to understand spin Hall effects. First, the physical meaning of the quantities $x^\mu$ and $p_\mu$ described by the gravitational spin Hall equations might not be very transparent. However, this becomes clear in the case of the Mathisson-Papapetrou equations, as the momentum $p_\mu$ and the spin tensor $S^{\alpha \beta}$ are derived directly from the energy-momentum tensor of the considered object~\cite[Eq. 3.12]{Harte_2022}, and the worldline $x^\mu$ has the physical meaning of the energy centroid as defined by the chosen observer with $4$-velocity $t^\alpha$. 

Another well-known result for the Mathisson-Papapetrou equations is that they admit conserved quantities in spacetimes with Killing vectors $\kappa^\alpha$. The conservation law is
\begin{equation}
    p_\alpha \kappa^\alpha + \frac{1}{2} S^{\alpha\beta} \nabla_\alpha \kappa_\beta = \mathrm{const}.
    \label{consLaw}
\end{equation}
Thus, the same conservation law also holds for the gravitational spin Hall equations.

The role of the timelike vector field $t^\alpha$ can also be better understood from the perspective of the Mathisson-Papapetrou equations. In this context, it is a priori clear that the dynamics of a localized wave packet can be equivalently described by different worldlines, and there is no preferred way of fixing a particular one. The spin supplementary condition \eqref{eq:ssc} helps us to define a worldline $x^\mu(\tau)$, representing the energy centroid of the wave packet relative to an observer with $4$-velocity $t^\alpha$. Thus, it is natural to ask how different observers would describe the same wave packet. We have already seen in Sec.~\ref{sec:wigner} that even in flat spacetime, observers related by a boost will generally assign different energy centroids for the same wave packet. The dependence of the gravitational spin Hall equations on the timelike vector field $t^\alpha$ was first observed in Ref.~\cite{GSHE2020}, where it has been shown that the equations are nontrivial even in flat spacetime. Furthermore, it has been shown that, by the dependence on $t^\alpha$, the gravitational spin Hall equations encode the relativistic Hall effect~\cite{Relativistic_Hall} or the Wigner translations~\cite{Stone2015}. In a more general context, the role of $t^\alpha$ has been analyzed in Ref.~\cite{Harte_2022}. There, it has been shown how different descriptions of the same wave packet, assigned by two different observers with $4$-velocities $t^\alpha$ and $\tilde{t}^\alpha$, can be related. In the context of the Mathisson-Papapetrou equations, a change of observer can be viewed as a change of spin supplementary condition, and each observer will describe the wave packet in terms of a different set of quantities:
\begin{subequations}
\begin{align}
    t^\alpha:& \qquad \{ x^\mu, p_\mu, S^{\alpha \beta} = \epsilon s \frac{\varepsilon^{\alpha \beta \mu \nu} p_\mu t_\nu}{p_\sigma t^\sigma} \}, \quad S^{\alpha \beta} t_\beta = 0, \\
    \tilde{t}^\alpha:& \qquad \{ \tilde{x}^\mu, \tilde{p}_\mu, \tilde{S}^{\alpha \beta} = \epsilon s \frac{\varepsilon^{\alpha \beta \mu \nu} \tilde{p}_\mu \tilde{t}_\nu}{\tilde{p}_\sigma \tilde{t}^\sigma} \}, \quad\tilde{S}^{\alpha \beta} \tilde{t}_\beta = 0.
\end{align}
\end{subequations}
Then, up to error terms of order $\epsilon^2$, these two sets of quantities are connected throgh the relations~\cite{Harte_2022}
\begin{subequations}
\begin{align}
    \tilde{x} &= \exp_{x} \xi^\alpha + \mathcal{O}(\epsilon^2), \\
    \tilde{p}_{ \alpha' } &= g^{\alpha}{}_{ \alpha' } p_\alpha + \mathcal{O}(\epsilon^2), \\
    \tilde{S}^{ \alpha' \beta' } &= g^{ \alpha' }{}_{ \alpha } g\indices{^{ \beta' }_{\beta}} ( S^{\alpha \beta} + 2 p^{[\alpha} \xi^{\beta]} ) + \mathcal{O}(\epsilon^2),
\end{align}
\end{subequations}
where $\exp$ is the exponential map on the tangent bundle and $g^{\alpha}{}_{ \alpha' }$ is the parallel transport operator from $x$ to $\tilde{x}$, along the geodesic with the tangent vector $\xi^\alpha$. This transformation is defined by the spacelike shift vector
\begin{equation}
    \xi^\alpha = \frac{ S^{\alpha \beta} \tilde{t}_{\beta} }{  p_\sigma \tilde{t}^{\sigma} } + \mathcal{O}( \epsilon^2).
    \label{xiGen1}
\end{equation}
Thus, the above equations relate any possible descriptions of the same wave packet that different observers might assign. Two different observers will generally assign different worldlines to the wave packet, and these will be related by the shift vector $\xi^\alpha$. Note that the shift vector always vanishes in regions of spacetime where $t^\alpha = \tilde{t}^\alpha$, and then the assigned worldlines coincide (see Ref.~\cite[Fig. 1]{Harte_2022}). Furthermore, the above transformation laws are exact in flat spacetime, in which case they coincide with the Wigner-Souriau translations discussed in Refs.~\cite[Eq.~3.7]{Duval_chiral_fermions} and~\cite[Eq.~2.7]{DUVAL2015} for chiral fermions.

\subsection{The optical metric -- recovering the spin Hall effect of light} \label{sec:optical_metric}

One of the initial arguments for the presence of a gravitational spin Hall effect was the analogy between electromagnetic waves propagating in optical media and electromagnetic waves propagating on curved spacetime. This has been discussed in Sec.~\ref{sec:analogue}. Now that we have derived the gravitational spin Hall equations independently of this analogy, we can also apply the results discussed in Sec.~\ref{sec:analogue} in reverse order. 

It has been shown in Ref.~\cite[Sec. VI]{Harte_2022} that the equations of motion describing the spin Hall effect of light in an optical medium with refractive index $n$, as given here in Eq. \eqref{eq:SHEL}, can be recovered from the gravitational spin Hall equations \eqref{eq:gshe_eq} by using the optical metric~\cite{Gordon,Synge1960,covariant_dielectric}
\begin{equation} \label{eq:optical_metric}
    g_{\alpha \beta} = \tilde{g}_{\alpha \beta} + (1 - n^{-2}) u_\alpha u_\beta.
\end{equation}
This is an effective metric that describes the propagation of light when a dielectric medium with refractive index $n$ and $4$-velocity $u^\alpha$ is present in a background spacetime with metric tensor $\tilde{g}_{\alpha \beta}$. To replicate the physical scenario in which the spin Hall effect of light has been derived and measured, we take the Minkowski background metric $\tilde{g}_{\alpha \beta} = \eta_{\alpha \beta}$ in inertial coordinates $(t, x, y, z)$, and a dielectric medium with refractive index $n(x,y,z)$. The $4$-velocity of the medium, as well as the $4$-velocity of the observer defining the centroid, are taken as $u^\alpha = t^\alpha = (1, 0, 0 , 0)$. Then, the gravitational spin Hall equations \eqref{eq:gshe_eq} reduce to the equations of motion that describe the spin Hall effect of light in an inhomogeneous optical medium, derived in Refs.~\cite{SHE_original, Onoda2006,Bliokh2004,Bliokh2004_1,Bliokh2008,Bliokh2009,Ruiz2015} and given here in Eq. \eqref{eq:SHEL}.

This result represents an important test for the gravitational spin Hall equations, as it provides a direct link between a theoretical prediction and a phenomenon that has been observed in several optical experiments. Furthermore, the covariant form of the gravitational spin Hall equations, as well as their connection with the Mathisson-Papapetrou equations, could help us improve our understanding of spin Hall effects in inhomogeneous media. One such aspect, which is evident now from the covariant equations \eqref{eq:gshe_eq} but not from Eq. \eqref{eq:SHEL}, is the role of the observer. Thus, the gravitational spin Hall equations can be viewed as a covariantly unified framework, which brings together the relativistic Hall effect and the spin Hall effect of light.

\section{Applications and astrophysical relevance} \label{sec:applications}

Spin Hall effects have been observed in several experiments in optics~\cite{Hosten2008,Bliokh2008,SOI_review} and condensed matter physics~\cite{originalSHE3,originalSHE4,SHE_review}, and are currently playing an important role in the development of a wide range of applications~\cite{SHEL_metrology,Wunderlich1801,Jungwirth2012,liu2017photonic,photonics2018,Optical_comm_2016,SHE_image_proc2019}. This represents a strong motivation to extend and study these effects in the presence of gravity, from both a theoretical and observational point of view. 

The gravitational spin Hall effect, as discussed in Sec.~\ref{sec:SHE_gravity}, is now based on a solid theoretical framework, with several consistency checks performed and with known results recovered in special cases. The gravitational spin Hall equations \eqref{eq:gshe_eq} describe the propagation of circularly polarized electromagnetic or gravitational wave packets, and it is natural to address the question of experimental observability at this point. 

In this section, we present some possible avenues that could lead to an experimental observation of the gravitational spin Hall effect. We start in Sec.~\ref{sec:lensing_gw} by reviewing the theoretical results obtained in Ref.~\cite{GSHE_lensing}, where the strong-field lensing of gravitational waves has been studied based on the gravitational spin Hall equations. In Sec.~\ref{sec:shadow}, we briefly discuss polarization-dependent effects on black hole shadows.

\subsection{Frequency and polarization-dependent lensing of gravitational waves} \label{sec:lensing_gw}

\begin{figure}[t]
    \centering
    \includegraphics[width=0.8\columnwidth]{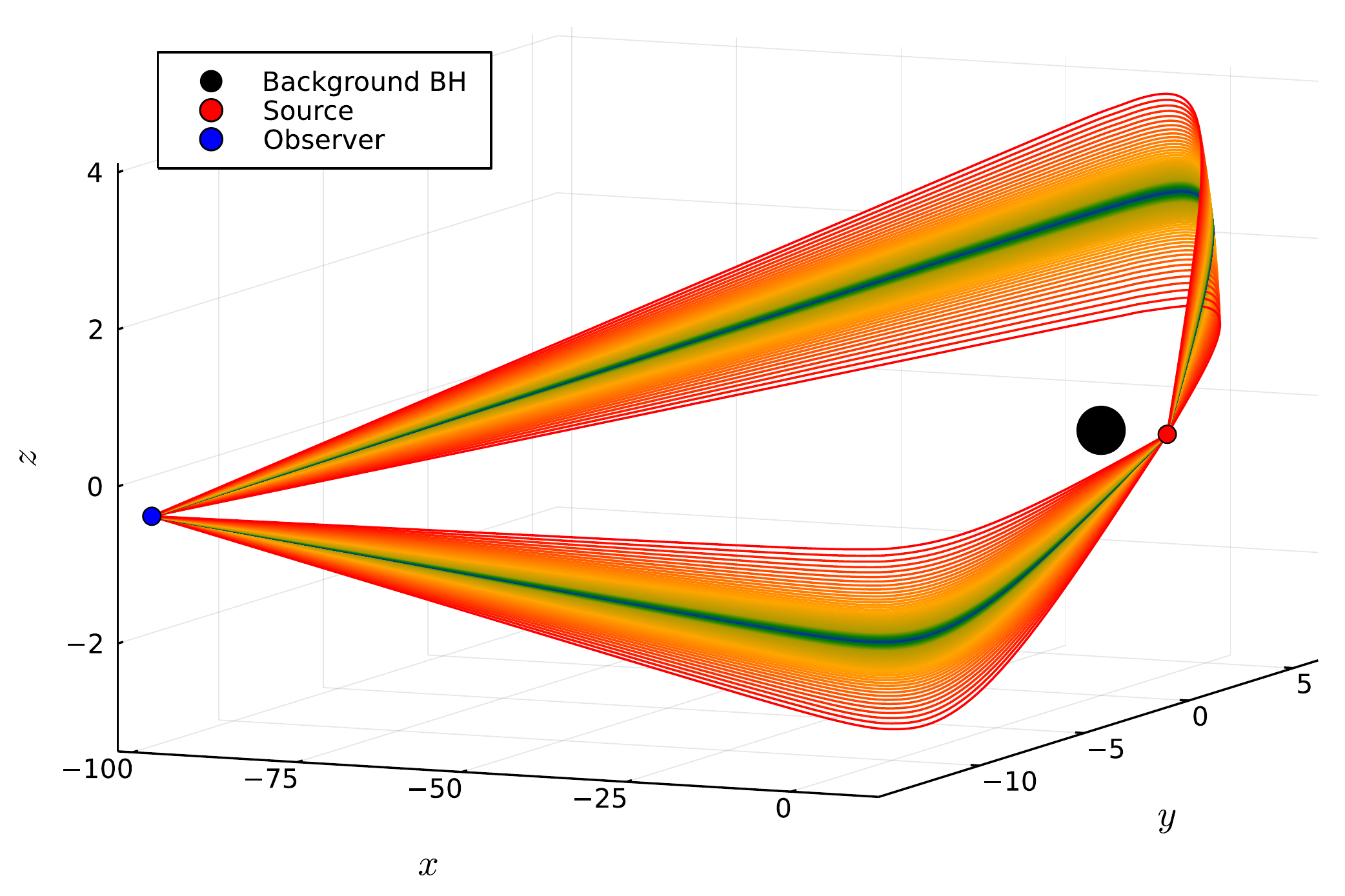}
    \caption{Gravitational spin Hall effect of strongly lensed gravitational waves originating from a hierarchical triple black hole system. Two small black holes merge in the vicinity of a much larger black hole which acts as a lens. The emitted gravitational waves are lensed and travel towards an observer along two bundles of connecting spin Hall trajectories (trajectories that loop several time around the black hole are ignored). Along each bundle, the rainbow colors depict the frequency dependence of the gravitational spin Hall trajectory, with geodesics recovered in the infinite frequency limit (blue trajectories). Each bundle contains two rainbows, one for each state of circular polarization ($s = \pm2$).  
    }
    \label{fig:example_trajectory}
\end{figure}

The magnitude of the gravitational spin Hall effect is proportional to the wavelength and to the strength of the gravitational field experienced by the wave packet. However, the wavelength cannot be too large compared to the characteristic length scales of the gravitational field, as this would violate the assumptions of the WKB approximation used in the derivation of the effect. This suggests that we should be looking for a significant effect in astrophysical systems where wave packets of large wavelengths are experiencing strong gravitational fields. 

Using current ground-based detectors, we can observe electromagnetic radio waves with maximum wavelengths on the order of $10$ meters (this is essentially limited by transmission through the ionosphere), but we can observe gravitational waves with wavelengths that are several orders of magnitude larger, typically on the order of $10^6$ meters~\cite{TheLIGOScientific:2014jea, Abbott:2016blz}. Therefore, gravitational waves are the better candidate for a possible observation of the gravitational spin Hall effect. 

A theoretical and numerical study of the gravitational spin Hall effect experienced by strongly lensed gravitational waves has been presented in Ref.~\cite{GSHE_lensing}. There, hierarchical triple black hole systems were considered, where two small black holes are merging and emitting gravitational waves in the vicinity of a third, much larger black hole, which acts as a lens. Thus, in this scenario the emitted gravitational waves are likely to experience gravitational lensing in the strong gravitational field of the large black hole, which could result in a significant gravitational spin Hall effect.

The general setup studied in Ref.~\cite{GSHE_lensing} is illustrated in Fig. \ref{fig:example_trajectory}. The hierarchical triple is modeled as a fixed background Kerr black hole, which acts as a lens, together with a static point source of gravitational waves, close to the Kerr black hole. Gravitational waves are measured by a static observer placed far away from the Kerr black hole. As long as the wavelengths of the emitted gravitational waves are small in comparison to the Schwarzschild radius of the background Kerr black hole, their propagation can be accurately described by the gravitational spin Hall equations \eqref{eq:gshe_eq}. 

Given a source-observer configuration as depicted in Fig. \ref{fig:example_trajectory}, the gravitational wave signal detected by the observer can be characterized by numerically calculating the connecting rays between the source and observer. Note that this is a much more difficult boundary value problem in space, while in Fig. \ref{fig:gshe_example} we illustrated an initial value problem for the spin Hall rays. There are generally two directly connecting null geodesics that do not loop around the black hole, together with infinitely many connecting null geodesics that loop several times around the black hole. We can safely restrict our attention to the directly connecting trajectories, as the ones that loop around the black hole will correspond to highly demagnified signals. The connecting null geodesics correspond to the infinite-frequency/zero-wavelength ($\epsilon = 0$) limit of the gravitational spin Hall equations \eqref{eq:gshe_eq}. As we increase the wavelength $\epsilon$, a ``double rainbow" bundle of connecting trajectories is formed along each connecting null geodesic. The ``rainbow" encodes the wavelength dependence of the gravitational spin Hall equations, and along each connecting null geodesic we get a double rainbow because for each wavelength $\epsilon$ we can have two states of circular polarization encoded by $s = \pm 2$. 

\begin{figure}
    \centering
    \includegraphics[width=0.9\columnwidth]{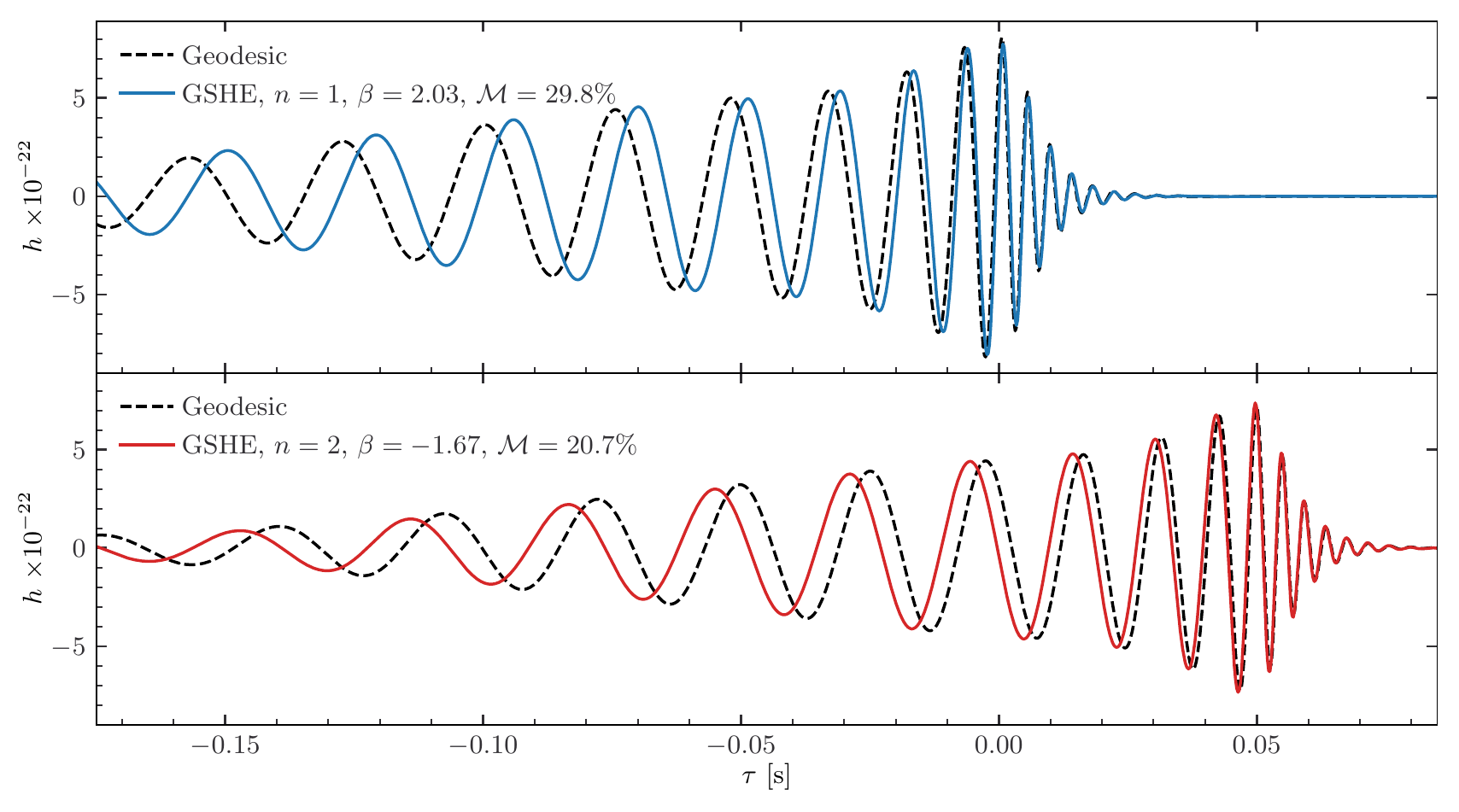}
    \caption{Imprints of the gravitational spin Hall effect on gravitational waveforms measured by a distant observer, corresponding to the configuration given in Fig.~\ref{fig:example_trajectory}. The dashed lines represent the gravitational waveform propagated between the source and observer along the null geodesics, while the solid lines represent the gravitational waveform propagated along the spin Hall trajectories. The upped and lower plots correspond to propagation of the signal along the two bundles labeled by $n=1,2$. Compared to the geodesic signal, the gravitational spin Hall effects manifests as a frequency- and polarization-dependent phase shift.}
    \label{fig:waveform}
\end{figure}

In Ref.~\cite{GSHE_lensing}, the gravitational spin Hall equations \eqref{eq:gshe_eq} were numerically integrated for several source-observer configurations. At the observer, the gravitational spin Hall effect manifests itself as a frequency- and polarization-dependent time of arrival of the gravitational waves. Within each bundle of connecting trajectories (labeled with $n$), the observer will measure the time delays $\Delta \tau^{(n)} (\epsilon, s)$ between the geodesic and the spin Hall rays, as well as the time delays $\Delta \tau_{R-L}^{(n)} (\epsilon)$ between the spin Hall rays of opposite circular polarization. The remarkable result obtained in Ref.~\cite{GSHE_lensing} is that, for all source-observer configurations that were tested, the time delays closely follow the power laws
\begin{equation}
    \Delta \tau^{(n)} (\epsilon, s) \propto \epsilon^2, \qquad \Delta \tau_{R-L}^{(n)} (\epsilon) \propto \epsilon^3.
\end{equation}
This means that for all source-observer configurations, the gravitational spin Hall-induced time delays can be characterized by a few numbers given by the proportionality factors of the above power laws. For the examples discussed in Ref.~\cite{GSHE_lensing}, the time delays $\Delta \tau^{(n)}$ are in the range $[10^{-6}, 10^{-3}]$ seconds, while the time delay between the different bundles of connecting trajectories was at least one order of magnitude larger ($\approx 50$ms for the example given in Figs. \ref{fig:example_trajectory} and \ref{fig:waveform}). Furthermore, $\Delta \tau^{(n)}$ is generally nonzero regardless of the spin of the Kerr black hole, whereas $\Delta \tau_{R-L}^{(n)}$ vanishes in Schwarzschild spacetimes. 

The gravitational spin Hall-induced time delays discussed above will modify the observed gravitational waveforms. For a gravitational waveform expressed in the frequency domain, any time delay can be simply included as a phase shift. Then, given an original unlensed waveform $\tilde{h}_{0}(f, s)$ in the frequency domain, the observed waveform including the gravitational spin Hall effect becomes
\begin{equation}\label{eq:GSHEwaveform_frequency}
    \tilde{h}_{\rm GSHE}(f, s) =
    \sum_n e^{-2\pi i f \Delta\tau^{(n)}(f, s)} \sqrt{\left|\mu^{(n)}(f, s)\right|} \tilde{h}_{0} (f, s).
\end{equation}
In the above equation, the sum runs over the different bundles of connecting trajectories and $\mu^{(n)}$ represents the magnification. Based on the source-observer configuration in Fig. \ref{fig:example_trajectory}, a gravitational waveform example is presented in Fig. \ref{fig:waveform}. We can clearly see that the gravitational spin Hall effect produces a significant modification of the waveform compared to the geodesically propagated signal. Given the frequency dependence of the time delay, we see a significant modification of the waveform in the low-frequency inspiral part, and the effect gradually fades away as we get closer to the high-frequency merger and ringdown parts of the signal. 

The difference between the gravitational spin Hall-lensed signal and the geodesically lensed signal can be quantified by calculating the mismatch $\mathcal{M}$~\cite[Eq. 2.28]{GSHE_lensing}. This represents a measure of distinguishability, and two waveforms are said to be distinguishable in a gravitational wave detector if the product of the mismatch with the squared signal-to-noise ratio is greater than $1$~\cite{Lindblom2008}. For the waveforms discussed in Ref.~\cite{GSHE_lensing}, as well as for the example presented here in Fig. \ref{fig:waveform}, the gravitational spin Hall effect typically produced a mismatch $\mathcal{M}$ in the range $1 \% - 30 \%$. Current ground-based gravitational wave detectors generally reach signal-to-noise ratios of the order $10$~\cite{abbott2021gwtc}, and this is expected to increase by at least one order of magnitude with future upgrades and space-based gravitational wave detectors. This suggests that the gravitational spin Hall effect on strongly lensed gravitational waves could be observed even with current ground-based detectors. This possibility, together with other astrophysical implications associated with this effect, is discussed in more detail in Ref.~\cite{GSHE_lensing}.

\subsection{Black hole shadows} \label{sec:shadow}

Black hole shadows represent the most extreme regime of gravitational lensing, where light rays passing close to the event horizon of a black hole are deflected so much that they loop several times around the black hole or can even become trapped. More precisely, the shadow of a black hole can be defined on the celestial sphere of an observer as the set of incoming directions on which no light from background sources can reach the observer. 

The Event Horizon Telescope has recently reported observations of the shadows of supermassive black holes M87~\cite{EHT2019} and Sagittarius A*~\cite{EHT_SGRA}. These results are opening a new avenue for probing Einstein's theory of general relativity in the regime of strong gravitational fields. 

Motivated by recent observations, there are a vast number of theoretical studies exploring the properties of black hole shadows~\cite{BHS1,BHS2,BHS4,bhs5,bhs6,bhs7,bhs8,bhs9,Shadow_deg1,Shadow_deg2,Lupsasca2020,PhysRevD.102.124004,PhysRevD.101.084020} (for recent reviews, see Refs.~\cite{bhs7,Perlick2021}). However, all these studies are based on the geometrical optics approximation that light rays follow null geodesics. Therefore, polarization-dependent effects are generally ignored, with the notable exception of Ref.~\cite{PhysRevD.101.084020}, where the evolution of the polarization vector is considered along null geodesics, but there are still no polarization-dependent corrections to the ray equations. 

The strong gravitational fields generally encountered in the study of black hole shadows suggest that the gravitational spin Hall effect of light could also have a significant contribution. However, in this case, the wavelengths are much smaller (for example, the Event Horizon Telescope observes radio waves with wavelengths of $1.3$ millimeters) than those considered for gravitational waves in the previous section. Thus, it is not clear whether the gravitational spin Hall effect would be observable in this scenario, at least not with the current resolution of the Event Horizon Telescope. Nevertheless, an in-depth theoretical study of the gravitational spin Hall effect on black hole shadows would be required in order to properly assess the size of the effect and its prospects towards observability. Such a study is not available at this time, but some preliminary results were mentioned in Ref.~\cite[Sec. 2.5.6]{Oanceathesis}. There, the gravitational spin Hall equations \eqref{eq:gshe_eq} were used to numerically determine the shadows of Schwarzschild and Kerr black holes. Qualitatively, these results indicate that the shadows of Schwarzschild black holes are unchanged, whereas there are frequency- and polarization-dependent deformations for the shadows of Kerr black holes. We hope that this topic will be further explored in the near future.

\section{Conclusions} \label{sec:conclusions}

We have presented the basic properties of spin Hall effects arising across different areas of physics. These have been observed in several experiments in optics and condensed matter physics and have led to a wide range of applications. On the basis of recent predictions, similar effects are expected to influence the dynamics of wave packets propagating in inhomogeneous gravitational fields. Gravitational spin Hall effects represent a natural extension to the realm of general relativity of the well-known spin Hall effects present in optics and condensed matter physics. In this sense, black holes in spacetime can be seen as analogues of impurities in semiconductors or inhomogeneities of optical media, and gravitational spin Hall effects can be used as a probe for the curvature inhomogeneities of spacetime. 

Gravitational spin Hall effects arise as a consequence of spin-orbit coupling, a mechanism which determines the mutual interaction between the external (average position and momentum) and internal (spin, polarization, or intrinsic angular momentum) degrees of freedom of wave packets propagating in curved spacetime. This effect has been predicted for electromagnetic and linearized gravitational waves, as well as for massive and massless Dirac fields. As a consequence, such wave packets generally follow frequency- and polarization-dependent trajectories when propagating in curved spacetimes. 

From an experimental point of view, the gravitational spin Hall effect has not been observed so far. However, recent theoretical and numerical predictions~\cite{GSHE_lensing} suggest that strongly lensed gravitational waves could carry a significant imprint of the gravitational spin Hall effect, observable even with current gravitational wave detectors. An experimental observation of the gravitational spin Hall effect could have significant implications for astrophysics and would represent a novel probe of general relativity in the strong-field regime.

\section*{Acknowledgements}

The authors are grateful to Richard Stiskalek for his help with some of the figures.

\section*{References}
\bibliographystyle{iopart-num-mod}
\bibliography{references}

\end{document}